%Paper: cond-mat/9209025
%From: gammel@rats.nosc.mil (Jan T. Gammel)
%Date: Fri, 18 Sep 92 16:17:11 PDT

%%% figures appended at end
%%%%%%%%%%%%%%   cut here for body %%%%%%%%%%%%

%%%%%% format

\overfullrule=0pt
\nopagenumbers
\magnification=1100
\voffset -1.2truecm
\vsize=26truecm
\hsize=17.2truecm
\def\bls#1{\baselineskip=#1truept %plus1pt minus1pt
}

\def\title#1{\vbox to 2.5truecm{\parindent=0pt{#1}\vfil}  }
\def\authors#1{\vbox to 4.6truecm{\parindent=0pt{#1}\vfil}  }
\def\header#1{
$ $

\noindent {\bf #1} }

%%%%%% abreviations

\def \hb {\hfill\break}

\def\lessapprox{\hbox{$\lower1pt\hbox{$<$}\above-1pt\raise1pt\hbox{$\sim$}$}}
\def\greaterapprox{\hbox{$\lower1pt\hbox{$>$}\above-1pt\raise1pt\hbox{$\sim$}$}}

%%%%%%%%%%%%%%%%%%
\bls{16}
$ $
\vskip 1.8truecm
%%%%%%%%%%%%%%%%%%
%\noindent Keywords: POLYACETYLENE, EXCITED STATES, PEIERLS-HUBBARD MODELS

\title{
\underbar{
RELAXED EXCITED STATE GEOMETRIES
}\hfil\break
\underbar{
IN A PEIERLS-HUBBARD MODEL FOR POLYENES
}
}
\authors{
{J.~Tinka GAMMEL$^{(a)}$} and D.K.~CAMPBELL$^{(b)}$\hb
$^{(a)}$ Code 573, Materials Research Branch,
NCCOSC RDT\&E Division (NRaD),
San Diego, CA 92152-5000, USA \hb
$^{(b)}$Dept.~of Physics, University of Illinois at Urbana-Champaign,
1110 W.~Green St., Urbana, IL 61801, USA
}
\header{ABSTRACT}

Using parameter values relevant to
polyacetylene and its finite polyene analogs, we determine,
via exact finite size diagonalization of a 1D Peierls-Hubbard model,
the relaxed geometry and optical absorption spectra
of several states of interest in optical experiments (the
singlet ($1^1A_g$) and triplet ($1^3B_u$) ground states,
the singlet one ($n^1B_u$) and two ($2^1A_g$) photon gaps,
and the triplet one ($n^3A_g$) photon gap)
in order to examine the question of how well this simple model
can really capture the behavior of this simple material.

\header{INTRODUCTION}

Recent experimental and theoretical work have
established for the whole class of exciting novel
reduced dimensional materials, including the high-temperature superconducting
copper oxides, ``heavy-fermion'' systems,  organic synthetic
metals including conducting polymers, organic superconductors,
and charge transfer salts,
and halogen-bridged transition-metal chains,
that it is essential to include {\it both} the electron-electron (e-e)
and electron-phonon (e-p)
interactions if we are to truly understand the
properties of these materials. By carrying through
detailed comparison of model calculations with real materials,
we hope to achieve our modest goal: to gain an enhanced
understanding of the nature of the competition
between these two basic interactions of condensed matter physics.
The hope of achieving our goal is greatest, of course,
when the material and model
are as simple as possible while still retaining the essence
of the competition between the e-e
and e-p interactions. Thus, studying a linear chain material of simple
atomic constituents
should be ideal. In fact, polyacetylene, or (CH)$_x$,
has previously been touted as the ``hydrogen atom of the
conducting polymers.''

No interpretation of
dynamics of finite polyenes and conducting polymers can be considered
complete unless it correctly captures the important role of the
triplet and other excited
states in these systems. Triplet states in (derivatives of)
finite polyenes are known to be of considerable importance in
the processes of vision [1] and of photosynthesis
[2], in that they control certain crucial
relaxation paths after electronic excitation. In conducting polymers,
triplets appear essential to the understanding of the nature of long-lived
excited states and specifically of the spectra observed
in photoinduced photoabsorption. From a theoretical perspective,
knowledge of excited states and their properties
can provide additional insight
into the relative importance of electron-electron and electron phonon
interactions and can help determine the parameters in the theoretical
models of these materials.

\header{MODEL}

In the context of conducting polymers, the one-dimensional Peierls-extended
Hubbard Hamiltonian [3] provides a theoretical
framework capable of treating both electron-phonon (e-p)
and electron-electron (e-e) interactions of arbitrary strengths.
We focus on the one-dimensional Peierls-extended Hubbard Hamiltonian (1-D PHH)
in the context of {\it trans}-polyacetylene ({\it trans}-$(CH)_x$).
The 1-D PHH is
$$
H~=~
\sum_\ell (-t_0+\alpha\delta_\ell)
\sum_\sigma (
c^\dagger_{\ell~\sigma} c_{\ell+1~\sigma}+c^\dagger_{\ell+1~\sigma}
c_{\ell~\sigma}) ~
     +~ U \sum_\ell n_{\ell\uparrow} n_{\ell\downarrow}~
     +~ V \sum_\ell n_\ell n_{\ell+1} ~
     +~ {1\over 2} K\sum_\ell (\delta_\ell-{\rm a}_0)^2
\eqno{(1)}
$$
where
the general features of this Hamiltonian have
been described elsewhere [3].
Our studies here are a continuation of the work
described in Ref.s [4], [5], and [6]
as to it's specific applicability to polyacetylene.
We now can incorporate additional excited state information.

Due to space limitations, we refer the reader to Ref.s [4] and [5]
for our determination of polyene parameters using ground state data
and to Ref.~[6] for initial studies of excited geometries using conventional
polyacetylene parameters.
We found that, since the bandwidth is partly due to e-e and partly e-p
interactions, the hopping integral is less than the conventional
$t_0$=2.5 eV value, but the parameter ratios remain roughly the
same. Our best fits were obtained with
$t_0$=1--1.5 eV, $\alpha$=3.5--4.5 eV/\AA, $U$=3--6 eV, $V$=0--2 eV (with the
fit roughly depending only on $U$$-$$2V$), and $K$=40--60 eV/\AA$^2$.
A typical absorption using parameters in this range showing
agreement with triplet and soliton absorptions
was shown in Fig.~3 of Ref.~[5].

\topinsert
\vglue 4.5truein
\bls{12}
\noindent
Fig.~1 (left). Relaxed geometry and total energy of the various
ground and excited states discussed in the text for
$U/t_0$=3 and $\alpha\delta/t_0$=0.1.
\vskip 12truept
\noindent
Fig.~2 (right). Same as Fig.~1, but for $\alpha\delta/t_0$=0.5.
For this electron-phonon coupling,
a higher lying $2^1A_g$ geometry can also be found.
\endinsert
\bls{16}

\header{EXCITED STATES}

Using the method for extracting excited states described in the appendix,
we use the Lancz\"os method to determine the energies and wavefunctions
of the ground state and several important excited states:
the lowest triplet state ($1^3B_u$), the second singlet
``$A_g$'' state ($2^1A_g$) which determines the
two-photon gap, and the ``optical states'' which determine the
singlet optical gap ($n^1B_u$) and triplet optical gap ($n^3A_g$) [7].
Initial studies using conventional polyacetylene parameters
and a fixed uniformly dimerized geometry
were reported in Ref.~[6]. Initial relaxed studies of only
the triplet geometry were also reported in that paper.
We have now obtained relaxed geometry information
for all these excited states.
Both conventional polyacetylene parameters and our new
set given above have roughly $U/t_0$=3
and $\alpha\delta/t_0$=0.1, where $\delta$
is the average dimerization in the ground state [8].
The relaxed geometries for this (dimensionless) parameter set
are shown in Fig.~1 for a 12-site system using open boundary
conditions. Rather than the raw distortion, we have plotted
the distortion order parameter $\pm(-1)^n\delta_n$ (with the sign
being chosen so that a short bond at the chain end is positive)
so that the deviation from uniform distortion is more apparent.
We also plot the total energy for each of these
states at each of these geometries to show the expected
luminescence shifts, as well as check that our
numerical procedure converged to the lowest energy configuration and
not a higher lying local minimum.
Unfortunately, for this small an electron-phonon
coupling, it is difficult to analyze the geometry dependence as
it is so weak. Thus, following Ref.~[9], we increase the electron-phonon
coupling until  $\alpha\delta/t_0$=0.5. The results are shown
in Fig.~2.
Note that the singlet gap at the relaxed geometry of the optical
state is approximately half its value at the ground state
geometry.
We also show in Fig.~3 the size dependence of the relaxed geometries
for these same parameters.
The state labeled ``$2^1A_g$" is higher in energy
than the state labeled ``$2^1A_g^*$"
on small systems ($N$=8,10)
but becomes lower in energy on larger systems ($N$=12).
Note that in the ground state geometry,
the $2^1A_g$ lies roughly twice as far above the ground state
as does the $1^3B_u$ state. This strongly hints that the
$2^1A_g$ state is in some sense composed of two triplets,
as suggested also by the work of Tavan and Schulten [10].  From
Fig.~3 we see that the  $1^3B_u$ state has the
lattice distortion appropriate to {\it two} neutral solitons,
and the $2^1A_g$ prefers a 4-soliton configuration at $N$=12.
Does this
mean that one should view the $2^1A_g$ state as composed of {\it four}
solitons? Such an interpretation was proposed in the renormalization
group studies of Hayden and Mele [11] but
is explicitly contradicted in the configuration interaction studies
[10] and (at least apparently) in combined
experimental/theoretical studies on octatetraene [12].
Unfortunately, it is difficult to determine a definitive
answer to this question (see Ref.s [3] and [6] for a
more thorough discussion).
At present, our data
on the {\it fully relaxed} geometry and energy of the $2^1A_g$ state
on the largest systems ($N$ = 12,14) are not sufficient to
determine the resolution of this question unambiguously. This
difficulty arises in part because the $2^1A_g$ state is not
the lowest singlet state of $A_g$ symmetry and thus must be extracted
from the Lancz\"os data with considerable care.
However, to date our Lancz\"os results are most consistent with
the ``4-soliton" viewpoint for long polyenes, and not inconsistent
with the experiment showing a different geometry for octatetraene.

\topinsert
\vglue 4truein
\bls{12}
\noindent
Fig.~3. Size dependence of the relaxed geometries
for the parameters of Fig.~2.
For $N$=8,10 the configuration labeled $2^1A_g^*$ is preferred,
but at $N$=12 the configuration labeled $2^1A_g$
becomes lower in energy.
\endinsert
\bls{16}

$ $

\underbar{\it Acknowledgements.}
This work would have been impossible without
the close collaboration of
E.Y.~Loh,~Jr., S.~Mazumdar and B.~Kohler.
We  also wish to thank
A.R.~Bishop, J.~Bronzan, R.~Friend, and V.~Vardeny, among others,
for many useful discussions.
JTG was supported by a NRC/NRaD
Research Associateship through a grant from ONR.
Computational support was provided by the ACL, CMS, and CNLS at LANL.

%%%%%%%%%%%%%%%%%%%%%% lanczos procedure %%%%%%%%%%%%%%%%%%%

\header{APPENDIX: NUMERICAL METHOD FOR OBTAINING EXCITED WAVEFUNCTIONS}

Our ``exact'' diagonalizations were performed using a standard
Lancz\"os algorithm
which in essence involves expressing the Hamiltonian in a
cleverly chosen basis, in which it is tridiagonal.
Only the electronic part of the Hamiltonian was treated
exactly; the adiabatic approximation was used for the phonons.
Our formulation of the standard
Lancz\"os procedure and method for extracting ground state
relaxed geometry and optical absorption information,
including our averaging technique for enhancing
the finite-size results,
has been previously described (Ref.~[13]).

While the standard Lancz\"os method is useful for projecting out
ground states, we are typically interested in several states
which are important for, $e.g.$, optical experiments:
the $1^1A_g$ (ground state), $2^1A_g$ (two-photon singlet gap),
$n^1B_u$ (singlet optical gap, $1^1B_u$ for small $U$),
$1^3B_u$ (triplet ground state), and $n^3A_g$
(triplet gap, $1^3A_g$ for small $U$).
While the standard Lancz\"os method
allows us to get the singlet and triplet (by working
in the $S_z$=1 subspace) ground state
wavefunctions and energies, it allows us to find only the energy
of the one and two photon gap states, and not the
wavefunctions of these excited states,
and hence we cannot investigate, $e.g.$, questions
of geometry relaxation upon excitation and corresponding luminescence
spectra. However, by slightly modifying the Hamiltonian and
procedure, we can indeed
obtain such information, though some care must be taken as
occasionally the numerical procedure fails and so one must test
that the resultant states are indeed eigenfunctions of the
original Hamiltonian with
the desired symmetry.
Once they have been found, evaluation of correlation
functions (such as the bond-charge, which determines
the self-consistent relaxed geometry) proceeds as for the ground state.
Though we describe here the method only for a single boundary
condition, the generalization to our "boundary condition
averaged" scheme, described in Ref.~[13]
is straightforward.

The first step [14] is to note that the starting Hamiltonian,
Eq.~1,
commutes with several symmetry (parity) operators, especially
if the distortion has some symmetry.
Assuming there are equal numbers of up and down spin
electrons and that the distortion
is symmetric about some bond (or site), call it the $n$-th one, then we can
add the following terms to the Hamiltonian without changing the
eigenfunctions:
$$
H_1 ~=~\lambda_F F~+~\lambda_R R~+~\lambda_S S^2~
$$
Where $S$ is the total spin, $R$ is the parity operator
which inverts about the $n$-th bond (site), and $F$ flips every spin.
When $H=H_0+H_1$ is applied to a wavefunction, one finds
$$\eqalign{
state~~~ &Energy \cr
^1A_g~~~~ & E_0(^1A_g)+\lambda_F+\lambda_R\cr
^1B_u~~~~ & E_0(^1B_u)+\lambda_F-\lambda_R \cr
^3B_u~~~~ & E_0(^3B_u)-\lambda_F-\lambda_R+2\lambda_S\cr
^3A_g~~~~ & E_0(^3A_g)-\lambda_F+\lambda_R+2\lambda_S \cr
}$$
Thus we can, $e.g.$, preferentially select the singlet over the triplet
ground state by setting $\lambda_F$ to a large number and performing
the standard Lancz\"os procedure.
One could also apply a projection operator $P$ which projects
out a given total spin. However, this particular operator is
($i$) computationally expensive and ($ii$) tends to cause the numerical
procedure to fail (through, $e.g.$, loss of orthogonality).
Thus we have not employed it. Further, raising or lowering the
energy based on the total spin tends to favor zero (singlet) and
{\it maximum} spin, rather than the singlet and triplet, and
flipping the spin has the same singlet/triplet selectivity, thus
we have only used the symmetries $F$ and $R$ to project out
eigenstates, and tested the resultant state to insure the
desired total spin.

To project out the $1^1B_u$ is a simple matter of adjusting the
$\lambda$ and finding the ``ground" state (of $H$, first allowed
$^1B_u$ for $H_0$) with the usual Lancz\"os
procedure. However,
for large $U$, the $1^1B_u$ is not the optical edge [7].
One can get around this problem by
a slight modification of the Lancz\"os procedure. If we assume the
ground state $|\phi_0\rangle$ is known, then after we have generated
a given basis vector $|\psi_j\rangle$ we also calculate
and store $\langle\phi_0|J|\psi_j\rangle$.
Once the storage limit of the computer is reached, instead
of selecting the eigenvector with lowest eigenvalue,
we select the one with lowest eigenvalue {\it and} non-zero
$\langle\phi_0|J|\phi_k\rangle$. (To do this we need only deal
with the small matrix which converts a Lancz\"os basis number to
a eigenvector number, thus this step does not add considerably
to the time or size of the calculation). We iterate this procedure
until the estimate of the $n^1B_u$ energy stops changing, as for the
ground state.
Note that this ($i$) only increases the memory requirement by
one state, and ($ii$) only adds one ``expensive" step per Lancz\"os
basis state.
Of course, the statement ``non-zero" above means
in practice larger than
some cutoff. If too small a cutoff is used, the $1^1B_u$, $2^1A_g$,
triplet ground state,  or other lower lying state not of
current interest, may be found instead of the desired optically
allowed state. We have found
a relatively large value of the cutoff
($\sim 10^{-2}$ in units of the matrix element  of the
optical edge) is required, though this is not as large as it seems,
as it is the square which is important for optical absorption.
It also means that, $e.g.$, for the triplet ground state
geometry, it is not the band edge, but rather the soliton
absorption, which is found (unless quite a large value for the
cutoff is used, which can be troublesome,
or an additional selection requirement that the
energy be ``near" the value for the uniform geometry,
assumed to have been previously calculated, is added).

At the end of this procedure for projecting out
a given excited state, it is important to
test that the resultant state  ($i$) is
an eigenstate of $H_0$, and ($ii$)  has the desired symmetry
and total spin,
to insure against loss of numerical accuracy.
(We have found using
$\lambda$ values on the order of the energy separation of interest
is usually successful.)
Obviously, while straightforward, this excited state
projection procedure cannot
be used blindly, and it is important to test not only that
the ``answer" is an eigenstate with correct symmetry and spin
(and non-zero $\langle\phi_0|J|\phi_k\rangle$
if an optically allowed state is desired)
but that comprehensible trends
as a function of system size or other parameter are followed,
before one can have full confidence that one has indeed calculated
what one set out to.

When there are unequal numbers of up and down spin electrons
or the distortion has no inversion symmetry,
the symmetry operators $F$ or $R$, respectively, cannot be used,
though the procedure described above (with obvious modifications)
can still be used to project out, $e.g.$, the wavefunction of the
first optically allowed state.
Also it is clear that one can obtain further excited state
information ($e.g.$, second optically allowed state)
using these and similar modifications of the basic
Lancz\"os procedure, though at the
cost of larger time and storage requirements and
(potentially fatal) loss of numerical accuracy.
We note that further excited state information,
such as the many-body
density of states, can also be obtained in a related but somewhat
different fashion which essentially involves recursively generating
the basis to fill in the necessary matrix element of the
correlation function of interest and which does not significantly
increase the storage requirements or decrease the numerical
accuracy, though it obviously increases the time
requirements [15].

\header{REFERENCES}
\parskip=0pt

\item{1}
R.S. Becker {\it et al.,}
{\it J. Chem Soc. Faraday Transactions 2,} {\bf 74} (1978) 2246.
\item{2}
J. Lafferty {\it et al.,}
{\it J. Chem Soc Faraday Transactions 2,} {\bf 73} (1976) 416.
\item{3}
A survey of the Peierls-extended Hubbard
model and its applications to conducting polymers is given in
D. Baeriswyl, D. K. Campbell, and S. Mazumdar,
in H. Kiess (ed.), {\it Conducting Polymers}, (Springer, 1991).
\item{4}
J.T.~Gammel, D.K.~Campbell, E.Y.~Loh,~Jr., S.~Mazumdar, and S.N.~Dixit,
{\it Mat. Res. Soc. Symp. Proc.,} {\bf 173} (1990) 419.
\item{5}
J.T.~Gammel, D.K.~Campbell, S.~Mazumdar, S.N.~Dixit, and E.Y.~Loh,~Jr.,
{\it Synth.~Metals}, {\bf 43} (1991) 3471.
\item{6}
D.K.~Campbell, J.T.~Gammel, H.Q.~Lin, and E.Y.~Loh,~Jr.,
{\it Synth.~Metals}, {\bf 49} (1992) in press.
\item{7}
At large $U$, the lowest optical transition is no longer
to the first state of
odd symmetry, $1^1B_u$, but to a higher lying $n^1B_u$ state,
as the energy of this optical state increases roughly
linearly with $U$.
Similarly, the triplet optical state has a cross-over and
also becomes a higher lying $n^3A_g$ state at large $U$.
\item{8}
At $N$=12, $\alpha\delta/t_0$=0.1, $U/t_0$=3.0
yields 1/$\pi\lambda$=$Kt_0/2\alpha^2$=2.4,
and $\alpha\delta/t_0$=0.5 yields 1/$\pi\lambda$=0.77.
\item{9}
S.N.~Dixit, D.~Guo and S.~Mazumdar,
{\it Phys. Rev., B} {\bf 43} (1991) 6781.
\item{10}
P. Tavan and K. Schulten,
{\it Phys. Rev., B} {\bf 36} (1987) 4337.
\item{11}
G. W. Hayden and E. J. Mele,
{\it Phys. Rev., B} {\bf 32} (1985) 6527;
{\bf 34} (1986) 5484.
\item{12}
M. Aoyagi, I. Ohmine, and B.E. Kohler,
{\it J. Phys. Chem.,} {\bf 94} (1990) 3922.
\item{13}
D.K. Campbell, J.T.~Gammel, and E.Y. Loh, Jr.,
{\it Phys. Rev., B} {\bf 42} (1990) 475;
J.T.~Gammel, D.K. Campbell, and E.Y. Loh, Jr., this conference.
\item{14}
We are grateful to J.~Bronzan for suggesting this approach.
\item{15}
H. R\"oder and R. LeSar, unpublished.
%%%%%%%%%%%%%%%%%%%%%%

\end